\documentstyle[aps,epsfig,twocolumn]{revtex}

\begin{document}
\title{Scaling Ion Trap Quantum Computation through Fast Quantum Gates}
\author{L.-M. Duan}
\address{FOCUS Center and MCTP, Department of Physics, University of Michigan, Ann
Arbor, MI 48109 \\
}

\maketitle

\begin{abstract}
We propose a method to achieve scalable quantum computation based on fast
quantum gates on an array of trapped ions, without the requirement of ion
shuttling. Conditional quantum gates are obtained for any neighboring ions
through spin-dependent acceleration of the ions from periodic photon kicks.
The gates are shown to be robust to influence of all the other ions in the
array and insensitive to the ions' temperature.

\end{abstract}

Trapped ions constitute one of the most promising systems for
implementation of quantum computation. Significant theoretical
and experimental advances have been achieved for this system
\cite{1,2,3,4,5,6,7,8}, and quantum gates have been demonstrated
at the level of a few ions \cite{5,6,7,8}. The current central
problem is to find methods to scale up this system for
larger-scale quantum computation \cite{9,10,11,12,13,14}. A
particularly promising approach to scalability is based on
shuttling ions in complex traps to different regions for storage
and for quantum gate operations \cite {9,10,11}. Interesting
initial experiments have been reported on separating, shuttling,
and sympathetic cooling of the ions \cite{15,16,17}.
Nevertheless, these experiments also indicate that fast
separation of the target ions is a challenging task \cite{15},
which limits the speed of any collective quantum gate in a
scalable structure.

In this paper, we propose a scaling method based on fast quantum gates on an
array of trapped ions. Very recently, Garcia-Ripoll, Zoller, and Cirac
proposed a remarkable scheme of two-bit quantum gates \cite{18}, which can
operator faster than the ion trap frequency, where the latter was thought
before as a speed limit of the ion gates. The original fast gate in \cite{18}
was designed for two ions, with in mind that one achieves scalability
through the ion shuttling. Here, we propose an efficient scaling method for
ion trap quantum computation based on the concept of fast quantum gates.
This scaling method provides an alternative scheme to achieve scalability,
avoiding a series of challenges associated with the ion shuttling.

The extension of fast quantum gates from two ions to a large
array of ions is actually quite challenging: even if the laser
pulses are shined only on two ions, all the ions in the array
affect each other through the long-range Column interactions, and
all the phonon modes need to be taken into account as the
motional sideband addressing is not possible with a fast gate. The
mutual strong influence between the ions normally sets a
significant obstacle to scalable quantum computation. However, we
show that as long as the gate speed is faster than the local ion
oscillation frequency (specified below), this unwanted influence
can be arbitrarily reduced with a remarkable method for noise
cancellation. Besides the proposal of an efficient scaling
method, we also give a different design of the fast quantum gates
with the use of only periodic laser pulses. Such periodic pulses
are significantly easier for experimental realization.

We consider an array of trapped ions, which could be in any
convenient geometry. For instance, Fig. 1a shows a possible
configuration of the array from multi-connected linear Paul traps.
The qubit is represented by two stationary internal states of the
ion, denoted as $\left| 0\right\rangle $ and $\left|
1\right\rangle $ in general. We assume that the distance $d$
between the neighboring ions is appreciably large ( $d>5\mu m$)
so that single-bit operations and measurements can be done with
no difficulty. The critical issue is then to implement
prototypical two-bit gates for any neighboring ions, which,
combined with simple single-bit gates, realize universal quantum
computation.

\begin{figure}[tb]
\epsfig{file=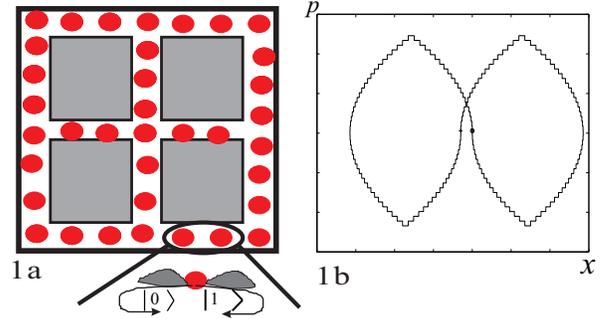,width=8cm} \caption{1a. A possible
geometry of the ion array for scalable quantum computing from
multi-connected linear Paul traps. Conditional quantum gates are
achieved for any neighboring ions though application of a
spin-dependent kicking force on these two ions. 1b. The typical
trajectory
of the kicked ion in phase space, where "$+$" denotes the initial position, "%
$\circ$" and "$\times$" denote the final position of the ion in $\left|
0\right\rangle $ or $\left| 1\right\rangle $ states, which in general is
different from the initial position as the ion has an initial velocity.}
\label{fig1}
\end{figure}

For the purpose of the two-bit gate on any neighboring ions, say, $i$ and $j$%
, we assume that one can mechanically accelerate each ion, with the
acceleration direction depending on its internal state $\left|
0\right\rangle $ or $\left| 1\right\rangle $ (as illustrated in Fig. 1a). So
the acceleration force is written as $F_{ij}=F\sum_{\alpha =i,j}\sigma
_{\alpha }^{z}$, where $\sigma _{\alpha }^{z}=\left| 1\right\rangle _{\alpha
}\left\langle 1\right| -\left| 0\right\rangle _{\alpha }\left\langle
0\right| $ is the Pauli operator. The state-dependent acceleration can be
realized, for instance, through coherent photon kicks from fast laser pulses
\cite{18}. A pair of short Raman pulses with the wave vectors ${\bf k}_{1}$%
and ${\bf k}_{2}$ applied on the ions $i$ and $j$ will give a
state-dependent momentum kick $\hbar \left( {\bf k}_{1}-{\bf k}_{2}\right)
\sigma _{\alpha }^{z}$ $\left( \alpha =i,j\right) $ to the ions for each $%
\pi $-rotation of the states $\left| 0\right\rangle _{\alpha }$ and $\left|
1\right\rangle _{\alpha }$. For convenience, we assume ${\bf k}_{1}\perp
{\bf k}_{2}$ and $\left| {\bf k}_{1}\right| \approx \left| {\bf k}%
_{2}\right| \equiv k_{c}=2\pi /\lambda _{c}$, where $\lambda _{c}$ is the
carrier wave length. The ${\bf k}_{1}$and ${\bf k}_{2}$ have a $45^{o}$
angle to the two-ion axis and we alternatively apply the Raman pulses in the
$\left( {\bf k}_{1},{\bf k}_{2}\right) $ and $\left( -{\bf k}_{1},-{\bf k}%
_{2}\right) $ directions. Each pulse sequence is periodic with a repetition
frequency $f_{r}/2$ much larger than the ion trap frequency, so the net
acceleration force on the ions is along the two-ion axis with the mean
magnitude $F=\sqrt{2}\hbar k_{c}f_{r}$. The momentum kicks from the periodic
laser pulses are actually discrete in time, but their effect is well
approximated by a continuous kicking force $F$ if more than ten pulses are
applied during each gate, as we have checked in our calculations.

To characterize this quantum gate, we need to take into account the Column
interactions between all the ions. The conventional approach is based on
decomposition into the canonical phonon modes. However, for a large number
of ions, this approach becomes very complicated even if one just wants to
find out all the phonon modes. Here, we develop a different theoretical
framework which is more appropriate for description of the fast gates on a
large ion array. We note that with fast laser pulses, the acceleration force
$F_{ij}$ can be significantly larger than the forces from the Column
interactions and the external trapping potentials. We therefore write the
total Hamiltonian $H$ into two parts $H=H_{0}+H_{1}$, where the dominant
part
\begin{equation}
H_{0}=-F\left( t\right) \sum_{\alpha =i,j}\sigma _{\alpha }^{z}x_{\alpha
}+\sum_{k}\frac{p_{k}^{2}}{2m}
\end{equation}
accounts for the kicking force $F\left( t\right) $ on the target ions $i,j$
and the kinetic energy summarized over all the ions $k$ (including $i,j$);
and the second part
\begin{equation}
H_{1}=\sum_{k,k^{\prime }}V_{kk^{\prime }}\left( \left| x_{k}-x_{k^{\prime
}}\right| \right) +\sum_{k}V_{T}\left( x_{k}\right)
\end{equation}
accounts for the Column interactions $V_{kk^{\prime }}\left( \left|
x_{k}-x_{k^{\prime }}\right| \right) $ between every pair of ions $%
k,k^{\prime }$ and the external trapping potentials\ $V_{T}\left(
x_{k}\right) $ on each ion $k$. In Eqs. (1) and (2), $x_{k}$ denotes the
displacement operator of the $k$th ion from its equilibrium position (so $%
\partial _{x_{k}}H_{1}\left( x_{k}\right) =0$ by definition) and $p_{k}$ is
the corresponding momentum operator \cite{sym}.

To first entangle and then disentangle the ions' internal and motional
states, we reverse the direction of the kicking force by two times as
depicted in Fig. 2a. So the force $F\left( t\right) $ in $H_{0}$ is time
dependent, with $F\left( t\right) =F$ for $0<t\leq t_{1}$ and $3t_{1}<t\leq
4t_{1}$, and $F\left( t\right) =-F$\ for $t_{1}<t\leq 3t_{1}$ (see Fig. 2a,
the value of $t_{1}$ will be specified later). Under these laser kicks, Fig.
1b shows the typical state-dependent trajectory of the $i$ or $j$ ion in
phase space under the ``free'' Hamiltonian $H_{0}$. Interestingly,
independent of the initial position or velocity of the ion, the wave packets
corresponding to the internal states $\left| 0\right\rangle $ and $\left|
1\right\rangle $ first split and then rejoin after a round trip. Under the
Hamiltonian $H_{0}$, the position operator $x_{k}$ of the $k$th ion evolves
as
\begin{equation}
x_{k}\left( t\right) =x_{k}+\left( t/m\right) p_{k}+s\left( t\right)
\sum_{\alpha =i,j}\delta _{k\alpha }\sigma _{\alpha }^{z},
\end{equation}
where $x_{k}$ and $p_{k}$ are the initial position and momentum operators,
and $s\left( t\right) =\int_{0}^{t}\int_{0}^{\tau _{1}}\left[ F\left( \tau
_{2}\right) /m\right] d\tau _{2}d\tau _{1}$.

\begin{figure}[tb]
\epsfig{file=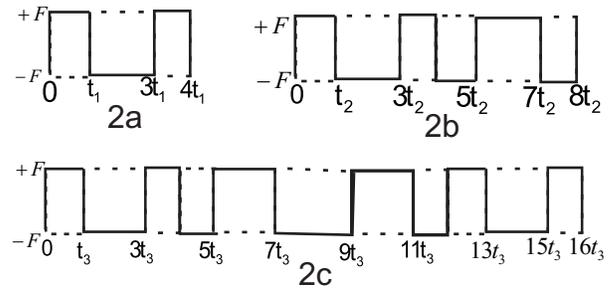,width=8cm} \caption{The sequence of the
spin dependent kicking force for the one-cycle (2a), two-cycle
(2b) and four-cycle (2c) schemes, with $t_{2}=t_{1}/2^{1/5}$ and $%
t_{3}=t_{1}/2^{2/5}$. The solid (dashed) curves are respectively
for the ions in $\left| 0\right\rangle $ or $\left|
1\right\rangle $ states.} \label{fig2}
\end{figure}

To investigate the internal state evolution of the ions, we transfer to the
interaction picture with respect to the ``free'' Hamiltonian $H_{0}$. The
Hamiltonian $H_{1}$ in the interaction picture becomes $H_{I}\left( t\right)
=U_{0}^{\dagger }\left( t\right) H_{1}\left( \left\{ x_{k}\right\} \right)
U_{0}\left( t\right) =H_{1}\left( \left\{ x_{k}\left( t\right) \right\}
\right) $, where $U_{0}\left( t\right) =\widehat{T}\left\{ \exp \left[
-i\int_{0}^{t}H_{0}\left( \tau \right) /\hbar d\tau \right] \right\} $ ($%
\widehat{T}\left\{ \cdots \right\} $ represents the time-ordered
integration) and $x_{k}\left( t\right) =U_{0}^{\dagger }\left( t\right)
x_{k}U_{0}\left( t\right) $ is expressed as Eq. (3). The evolution operator
in the interaction picture is given by $U_{I}\left( t\right) =\widehat{T}%
\left\{ \exp \left[ -i\int_{0}^{t}H_{I}\left( \tau \right) /\hbar d\tau %
\right] \right\} $. Note that at the end of the gate ($t=4t_{1}$), the $%
U_{0}\left( t\right) $ becomes an identity operators for the internal
dynamics (as $s\left( 4t_{1}\right) =0$ in Eq. (3)), therefore the gate
operation is wholly determined by the evolution operator $U_{I}\left(
t\right) $ at time $t=4t_{1}$.

To have an expression of $U_{I}\left( t\right) $, we note that there are
three different length scales in our problem. They are the distance between
the neighboring ions denoted as $d$, the length scale of the conditional
displacement $s\left( t\right) $ denoted as $\overline{s}$, and the
magnitudes of $x_{k}$ and $\left( t/m\right) p_{k}$ which are estimated by
the ion oscillation length and denoted as $\xi $. For fast quantum gates,
typically we have $d\gg \overline{s}\gg \xi $. Therefore, to the lowest ($0$%
th) order of the parameter $\xi /\overline{s}$, the coordinates $x_{\alpha
}\left( t\right) $ $\left( \alpha =i,j\right) $ are approximated by $s\left(
t\right) \sigma _{\alpha }^{z}$. In this limit, at time $t=4t_{1}$, the
internal and external dynamics of $U_{I}\left( 4t_{1}\right) $ become
disentangled, and the internal state evolution is described by
\begin{equation}
U_{\text{in}}=\exp \left[ -i\left( V_{ij}^{\prime \prime }/\hbar \right)
\int_{0}^{4t_{1}}s^{2}\left( t\right) dt\sigma _{i}^{z}\sigma _{j}^{z}\right]
,
\end{equation}
where the derivative of the Coulomb potential $V_{ij}^{\prime \prime
}=\partial _{x_{i}}\partial _{x_{j}}V_{ij}\left( \left| x_{i}-x_{j}\right|
\right) \simeq 2\hbar c/\left( 137d^{3}\right) $ ($c$ is the light
velocity). In writing Eq. (4), we have expanded the interaction Hamiltonian $%
H_{I}\left( t\right) $ to the second order of the small parameter $\overline{%
s}/d$ (the harmonic approximation). Equation (4) represents an ideal
controlled phase flip gate on the neighboring ions $i,j$ if $\left(
V_{ij}^{\prime \prime }/h\right) \int_{0}^{4t_{1}}s^{2}\left( t\right)
dt=\pi /4$, which determines the gate time
\begin{equation}
T_{g}=4t_{1}\simeq 7.09\left[ f_{r}^{2}v_{r}^{2}c/d^{3}\right] ^{-1/5}.
\end{equation}
In deriving Eq. (5), we have substituted the expression of the kicking force
$F$ into $s\left( t\right) $, and $v_{r}=\hbar k_{c}/m$ denotes the atom
recoil velocity. Figure (3a) shows the gate time $T_{g}$ as a function of
the repetition frequency $f_{r}$ of the laser pulses for the $^{111}Cd^{+}$,
$^{40}Ca^{+}$ and $^{9}Be^{+}$ ions. The gate speed ($1/T_{g}$) increases
with the kicking frequency as $f_{r}^{2/5}$ and decreases with the
neighboring-ion distance as $d^{-3/5}$. The total number of the laser pulses
for each gate is given by $N=f_{r}T_{g}$, which increases with the gate
speed as $\left( 1/T_{g}\right) ^{3/2}$. We can also expand the Hamiltonian $%
H_{I}\left( t\right) $ to higher orders of the small parameter $\overline{s}%
/d$ (beyond the harmonic approximation), which give some tiny correction to
the internal dynamics $U_{in}$ \cite{note2}. However, as long as we are in
the lowest order of the parameter $\xi /\overline{s}$, the internal and
external dynamics become disentangled after the pulse sequence, and there is
no intrinsic noise to the gate operation.

\begin{figure}[tb]
\epsfig{file=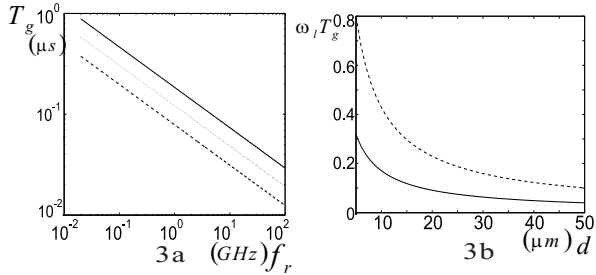,width=8cm} \caption{3a. The gate time
$T_{g}$ for the controlled phase flip operation is shown as a
function of the repetition frequency of the kicking laser pulses
for the $^{111}Cd^{+}$ (solid curve), $^{40}Ca^{+}$ (dotted
curve),
and $^{9}Be^{+}$ (dashed curve) ions, where the neighboring ions' distance $%
d $ is taken as $d= 10$ $\protect\mu $m, and the carrier wave length for the
$^{111}Cd^{+}$, $^{40}Ca^{+}$, and $^{9}Be^{+}$ ions are assumed to be $215$
nm, $393$ nm, and $313$ nm, respectively. 2b. The scaling parameter $\protect%
\omega _{l}T_{g}$ in the gate fidelity is shown as a function of
the neighboring ions' distance $d$ for $^{111}Cd^{+}$ with the
pulse repetition frequency $f_r=1$ GHz (dashed curve)and $f_r=10$
GHz (solid curve), respectively. The curve is well fit by the
scaling $\protect\omega _{l}T_{g}\propto d^(-9/10)$.} \label{fig3}
\end{figure}

We now examine the influence of the ions' oscillations by expanding the
interaction Hamiltonian $H_{I}\left( t\right) $ to the next ($1$st) order of
$\xi /\overline{s}$, which gives some intrinsic noise to the above
conditional quantum gate after the ions' motional states are traced. This
intrinsic noise comes from two respects: (i) The internal and motional
states of the $i$ and $j$ ions do not become fully disentangled after the
pulse sequence; (ii) Due to the mutual Coulomb interactions, the
state-dependent trajectory of the $i,j$ ions yields a state-dependent force
on all the other ions, which entangles the gate operation with the motional
states of all the ions in the array. When the Hamiltonian $H_{I}\left(
t\right) $ is expanded to the next-order of $\xi /\overline{s}$, the
evolution operator $U_{I}\left( 4t_{1}\right) $ after the pulse sequence is
decomposed as a product of the dominant part given by $U_{in}$ in Eq. (4),
and the noise part $U_{ns}$ given by
\begin{equation}
U_{ns}=\exp \left[ -i\sum_{\alpha =i,j;k}\sigma _{\alpha }^{z}\left( \beta
_{\alpha k}x_{k}+\gamma _{\alpha k}p_{k}\right) \right] .
\end{equation}
The coefficients $\beta _{\alpha k}=\left( 1-\delta _{\alpha k}/2\right)
H_{1\alpha k}^{\prime \prime }\int_{0}^{4t_{1}}s\left( t\right) dt$, $\gamma
_{\alpha k}=\left( 1-\delta _{\alpha k}/2\right) H_{1\alpha k}^{\prime
\prime }\int_{0}^{4t_{1}}\left( t/m\right) s\left( t\right) dt,$ and the
factor $H_{1\alpha k}^{\prime \prime }\equiv \partial _{x_{\alpha }}\partial
_{x_{\kappa }}H_{1}$ denotes the potential derivative at the ions'
equilibrium position. The operator $U_{ns}$ entangles the internal state of
the target ions with the motional states of all the ions in the array, and
introduces a gate infidelity after we take trace over the motional states.
To give a quantitative estimate of this intrinsic gate infidelity, we take
the following simplifications: (i) We denote $\omega _{L}=\sqrt{%
H_{1kk}^{\prime \prime }/m}$ as the local oscillation frequency of the ion $k
$ (with all the other ions fixed on their equilibrium positions) and assume
that this frequency is roughly the same for all the ions ($\omega _{L}$ is $k
$-independent); (ii) The states of the local oscillation modes $\left(
x_{k},p_{k}\right) $ are assumed to be thermal with the mean phonon number
estimated by $\overline{n}=k_{B}T_{t}/\left( \hbar \omega _{l}\right) $,
where $T_{t}$ is the temperature; (iii) The gate fidelity $F_{g}$ is defined
as $F_{g}=\left\langle \Psi _{0}\right| U_{in}^{\dagger }\rho
_{r}U_{in}\left| \Psi _{0}\right\rangle $, where $\left| \Psi
_{0}\right\rangle $ is the initial state of the $i,j$ ions, taken to be $%
\left( \left| 0\right\rangle _{i}+\left| 1\right\rangle _{i}\right) \otimes
\left( \left| 0\right\rangle _{j}+\left| 1\right\rangle _{j}\right) /2$ as
an example, and the final reduced internal state $\rho _{r}=Tr_{m}\left[
U_{ns}U_{in}\left| \Psi _{0}\right\rangle \left\langle \Psi _{0}\right|
U_{in}^{\dagger }U_{ns}^{\dagger }\right] $ is obtained by taking trace over
the motional states; (iv) We define the parameters $\xi ,\overline{s}$
quantitatively as $\xi =\sqrt{\hbar /\left( m\omega _{L}\right) }$ and $%
\overline{s}=\left[ \int_{0}^{4t_{1}}s^{2}\left( t\right) dt\right] /\left[
\int_{0}^{4t_{1}}s\left( t\right) dt\right] $. With these reasonable
simplifications, we find that the gate fidelity $F_{g}$ is estimated by $%
F_{g}\approx \exp \left[ -c_{1}\left( \overline{n}+1/2\right) \left( \xi /%
\overline{s}\right) ^{2}\right] ,$where the dimensionless constant $c_{1}$
is defined as $c_{1}=\pi ^{2}\left[ \left( H_{1ii}^{\prime \prime
}/2H_{1ij}^{\prime \prime }\right) ^{2}+\sum_{k\neq i}\left( H_{1ik}^{\prime
\prime }/H_{1ij}^{\prime \prime }\right) ^{2}\right] /4$. With the
definitions of $\xi ,\overline{s}$ and Eq. (5) for $T_{g}$, we can also
express the gate fidelity as
\begin{equation}
F_{g}\approx \exp \left[ -c_{2}\left( \overline{n}+1/2\right)
\omega _{L}T_{g}\right].
\end{equation}
The dimensionless constant $c_{2}\simeq 0.83c_{1}\left(
H_{1ij}^{\prime \prime }/H_{1ii}^{\prime \prime }\right) $. To
estimate the typical values of $c_{1}$ and $c_{2}$, in
calculating $H_{1ii}^{\prime \prime }$ we take into account only
the contributions from the intrinsic Coulomb interactions
(neglecting the contribution from the external trapping
potentials), and find that $c_{1}\approx 8.6$ $\left( 12.0\right) $ and $%
c_{2}\approx 3.0$ ($3.0$) respectively for a $1$ ($2$)-dimensional infinite
ion array. Note that the values of $c_{1}$ and $c_{2}$ are well bounded from
above even if we summarize over an infinite ion array, resulting form the
fact that the noise comes from an effective mutual dipole interaction
between the ions (described by the second derivatives of $H_{1}$), which
falls off very rapidly with the ions' distance.

The critical parameter $\omega _{L}T_{g}$ is shown in Fig. 3b as a function
of the distance $d$ between the neighboring ions with the laser repetition
frequency $f_{r}=1$ ($10$) GHz. Although this parameter scales down with
increase of the ions' distance, its value is not tiny in the typical
parameter region, and it is ineffective to reduce this parameter by further
increase of the laser repetition frequency due to the slow scaling $\omega
_{L}T_{g}\sim f_{r}^{-2/5}$. A significant intrinsic infidelity for the
quantum gate seems to be unavoidable if we substitute the typical value of $%
\omega _{l}T_{g}$ into Eq. (7). Fortunately, this is not the case as there
is an elegant way to greatly reduce this noise. In the above analyses, we
applied one cycle of the kicking force (shown in Fig. 2a) which pushes the
ions to the left (right) side if they are in the $\left| 0\right\rangle $ ($%
\left| 1\right\rangle $) state (see Fig. 1). We can improve the scheme by
using a two-cycle force as shown in Fig. 2b. Each cycle of the evolution
contributes the same amount of conditional phase which accumulated for a
conditional phase flip gate (so in Fig. 2 we should choose $%
t_{2}=t_{1}/2^{1/5}=T_{g}/4/2^{0.2}$). However, for these two cycles, the
ions are pushed to the reverse directions, and the coefficients $\beta
_{\alpha k}$ and $\gamma _{\alpha k}$ in the noise operator $U_{ns}$ (6)
thus have a reverse sign. Due to this sign flip, the noise effects from
these two cycles exactly cancel with each other! This noise cancellation is
perfect up to the $1$st order of the parameter $\xi /\overline{s}$. To
estimate the residue noise with a two-cycle force, we take into account all
the higher order contributions which basically rotate the operators $x_{k}$
and $p_{k}$ in Eq. (6) with a rotational angle $\theta \approx 4\omega
_{L}t_{2}\approx 0.87\omega _{L}T_{g}$ for two different cycles, so after a
partial cancellation of the noise, the accumulated coefficients $\beta
_{\alpha k}$ and $\gamma _{\alpha k}$ in the noise operator $U_{ns}$ are
reduced effectively by a factor of $1-e^{i\theta }\approx i\theta $, and the
final gate fidelity $F_{g2}$ for the two-cycle scheme thus has the form $%
F_{g2}\approx \exp \left[ -0.87^{2}c_{2}\left( \overline{n}+1/2\right)
\left( \omega _{L}T_{g}\right) ^{3}\right] $. We can further reduce the
noise by using more cycles of the kicking force to get noise cancellation up
to a higher order of the rotation angle $\theta $. For instance, Fig. 3c
shows a four-cycle kicking force, which reduces the accumulated noise
coefficients $\beta _{\alpha k}$ and $\gamma _{\alpha k}$ by a factor of $%
2\left( \omega _{L}T_{g}\right) ^{2}\times 4^{-2/5}$, and the
corresponding gate infidelity $F_{g4}$ is given by
\begin{equation}
F_{g4}\approx \exp \left[ -1.32c_{2}\left(
\overline{n}+1/2\right) \left( \omega _{L}T_{g}\right) ^{5}\right]
.
\end{equation}
With such a improvement, the gate infidelity $\delta F=1-F_{g4}$ has been
tiny in the typical parameter region. For instance, $\delta F\approx 0.006\%$
with the laser repetition frequency $f_{r}\approx 1$ GHz, the neighboring
ions' distance $d\approx 50$ $\mu $m, and thermal phonon number $\overline{n}%
\sim 1$.

Summing up, we have proposed a scheme to achieve scalable quantum
computation with trapped ions based on control from fast laser
pulses. This scheme has the following distinctive features: (i)
As the ion shuttling is not required in this scheme, one can use
an ion array in any convenient geometry from any types of ion
traps; (ii) The scheme is insensitive to the temperature of the
ions, and requires no challenging cooling of the ion crystal to
attain the Lamb-Dicke limit. The temperature $T_{t}$ affects the
gate fidelity through the mean phonon number
$\overline{n}=k_{B}T_{t}/\left( \hbar \omega _{L}\right) $ in Eq.
(8), and this influence is pretty weak. For instance, even with a
hundred of phonon excitations ($\overline{n}\sim 100$) right
after the Doppler cooling, the gate infidelity is still well
below $1\%$. (iii) The conditional gates are very fast even if we
have a pretty large distance between the neighboring ions which
allows easy separate addressing; (iv) The whole computation is
also fast as the slow step of separating the ions is not required
for achieving scalability in this scheme. The basic experimental
challenge is to well control the fast laser pulses to induce
series of coherent photon kicks. Periodic laser pulses with a
repetition frequency varying from hundreds of MHz to few THz have
been reported in many experiments \cite{book}. These impressive
achievements and rapid progress in control of short laser pulses,
combined with the ion trap technology, indicate realistic
prospects for building scalable ion trap quantum computers based
on this approach.

I thank Chris Monroe for helpful discussions and his valuable
suggestions. This work was supported by the Michigan start-up fund
and the FOCUS seed funding.


\begin{references}
\bibitem{1}  C. Monroe, {\it Nature} {\bf 416}, 238 (2002).

\bibitem{2}  J. I. Cirac and P. Zoller, Phys. Rev. Lett. 74, 4091 (1995).

\bibitem{3}  A. Sorenson and K. Molmer, Phys. Rev. Lett. 82, 1971 (1999);
Phys. Rev. A 62, 022311 (2000).

\bibitem{4}  G. J. Milburn, S. Schneider, and D. F. V. James, Fortschr.
Phys. 48, 801 (2000).

\bibitem{5}  C. Monroe et al., Phys. rev. Lett. 75, 4714 (1995).

\bibitem{6}  C. A. Sackett et al., Nature 404, 256 (2000).

\bibitem{7}  D. Liebfried et al., Nature 422, 412 (2003).

\bibitem{8}  F. Schmidt-Kaler et al., Nature 422, 408 (2003).

\bibitem{9}  D. J. Wineland et al., J. Res. Natl. Inst. Stand. Technol. 103,
259 (1998).

\bibitem{10}  J. I. Cirac and P. Zoller, Nature 404, 579 (2000).

\bibitem{11}  D. Kielpinksi, C. Monroe, and D. J. Wineland, Nature (London)
417, 709 (2002).

\bibitem{12}  A. B. Mundt et al., Phys. Rev. Lett. 89, 103001 (2002).

\bibitem{13}  B. B. Blinov, D. L. Moehring, L.-M. Duan. C. Monroe, submitted.

\bibitem{14}  L.-M. Duan, B. B. Blinov, D. L. Moehring, C. Monroe,
quant-ph/0401020.

\bibitem{15}  M. A. Rowe et al., Quantum Inf. Comput. 2, 257 (2002).

\bibitem{16}  B.B. Blinov {\it et al.}, Phys. Rev. A 65, 040304(R) (2002).

\bibitem{17}  M. D. Barrett et al., quant-ph/0307088.

\bibitem{18}  J. J. Garcia-Ripoll, P. Zoller, and J. I. Cirac, Phys. Rev.
Lett. 91, 157901 (2003).

\bibitem{sym}  We have written the operators $x_{k}$ and $p_{k}$ as scalars
for simplicity of symbols. It is straightforward to write them as
vectors to account for the higher dimensional case.

\bibitem{note2}  Keeping the expansion to the third order of the parameter $%
\overline{s}/d$, we actually get some additional single-bit rotations on the
ions $i$ and $j$ with the rotational angle typically about a hundredth of $%
\pi $. These small single-bit rotations can be compensated afterwards if
needed.

\bibitem{book}  G. P. Agrawal, Nonlinear Fiber Optics (Academic Press)
(2001).
\end{references}
\end{document}